%% file: spectrum-2014-12-11_jfm.tex
\documentclass{jfm}

\usepackage{graphicx}
\usepackage{natbib}
\usepackage{amsmath}
\usepackage[export]{adjustbox}

\ifCUPmtlplainloaded \else
  \checkfont{eurm10}
  \iffontfound
    \IfFileExists{upmath.sty}
      {\typeout{^^JFound AMS Euler Roman fonts on the system,
                   using the 'upmath' package.^^J}%
       \usepackage{upmath}}
      {\typeout{^^JFound AMS Euler Roman fonts on the system, but you
                   dont seem to have the}%
       \typeout{'upmath' package installed. JFM.cls can take advantage
                 of these fonts,^^Jif you use 'upmath' package.^^J}%
       \providecommand\upi{\pi}%
       \providecommand\ui{i}%
      }
  \else
    \providecommand\upi{\pi}%
    \providecommand\ui{i}%
  \fi
\fi

\ifCUPmtlplainloaded \else
  \checkfont{msam10}
  \iffontfound
    \IfFileExists{amssymb.sty}
      {\typeout{^^JFound AMS Symbol fonts on the system, using the
                'amssymb' package.^^J}%
       \usepackage{amssymb}%
       \let\le=\leqslant  \let\leq=\leqslant
         \let\geq=\geqslant
      }{}
  \fi
\fi

\ifCUPmtlplainloaded \else
  \IfFileExists{amsbsy.sty}
    {\typeout{^^JFound the 'amsbsy' package on the system, using it.^^J}%
     \usepackage{amsbsy}}
    {\providecommand\boldsymbol[1]{\mbox{\boldmath $##1$}}}
\fi

\newsavebox{\astrutbox}
\sbox{\astrutbox}{\rule[-5pt]{0pt}{20pt}}

\input{macros.tex}

\title[Control of the small-scale energy balance]{Direct control of the small-scale energy balance in 2D fluid dynamics}
\author[J. Frank, B. Leimkuhler, K.W. Myerscough]%
{Jason Frank$^1$%
  \thanks{Email address for correspondence: j.e.frank@uu.nl},\ns
Benedict Leimkuhler$^2$%
  \thanks{Email address for correspondence: b.leimkuhler@ed.ac.uk},\ns\break
and Keith W. Myerscough$^3$%
  \thanks{Email address for correspondence: k.w.myerscough@cwi.nl}}

\affiliation{$^1$Mathematical Institute, Utrecht University, P.O. Box 80010, 3508 TA Utrecht, the Netherlands\\[\affilskip]
$^2$School of Mathematics and Maxwell Institute for Mathematical Sciences, University of Edinburgh, James Clerk Maxwell Building, Kings Buildings Edinburgh EH9 3JZ, United Kingdom\\[\affilskip]
$^3$Centrum Wiskunde \& Informatica, P.O. Box 94079, 1090 GB Amsterdam, the Netherlands}

\pubyear{?}
\volume{?}
\pagerange{?--?}
\date{December 18, 2014}
\begin{document}

\maketitle

\begin{abstract}
We explore the direct modification of the pseudo-spectral truncation of 2D, incompressible fluid dynamics to maintain a prescribed kinetic energy spectrum. The method provides a means of simulating fluid states with defined spectral properties, for the purpose of matching simulation statistics to given information, arising from observations, theoretical prediction or high fidelity simulation. In the scheme outlined here, Nos\'{e}-Hoover thermostats, commonly used in molecular dynamics, are introduced as feedback controls applied to energy shells of the Fourier-discretized Navier-Stokes equations.    As we demonstrate in numerical experiments, the dynamical properties (quantified using autocorrelation functions) are only modestly perturbed by our device, while ensemble dispersion is significantly enhanced in comparison with simulations of a corresponding truncation incorporating hyperviscosity.   
\end{abstract}

\section{Introduction}
In fluid dynamics applications such as ensemble weather prediction and climate simulation, when the time scales of interest are long compared to the Lyapunov time, the goal of simulations is to accurately sample an evolving probability density function of the solution.   In general both the intermediate and steady-state distributions are unknown, being the consequence of forcing and dissipation introduced at various scales.   

Numerical truncation errors are often treated as random variables \citep{Evensen:09} with a known  (e.g.~zero-mean normal) distribution.
Recent improvements in the understanding of numerical methods, specifically the development of backward error analysis \citep{HaLuWa:06,Leimkuhler:04}, allow  interpretation of the numerical solution as the exact solution of a modified system of equations.  The modified equations typically admit their own (modified) invariant measure, and numerical truncation errors therefore bias the statistics obtained in simulation according to this altered statistical distribution.  Thus numerical methods imply structural bias due to numerical truncation, even when the continuum model is complete. 


As a precursor to accurately sampling an \emph{evolving} measure, it would seem essential that the numerical method accurately sample the \emph{stationary} invariant measure in the absence of forcing and dissipation, to allow correct response of the system to perturbations from equilibrium; however, even this requirement is typically not fulfilled, as has been observed in numerical investigations of simple two-dimensional ideal fluids models.  In \citep{DuFr:07} it was shown that the equilibrium statistical mechanics of finite difference discretizations of quasigeostrophic vorticity flow over topography are sensitive to the preservation of kinetic energy and (quadratic) enstrophy.  Even in the idealized setting of unforced, inviscid 2D flow, a correct sampling of non-Gaussian statistics in the Miller-Robert-Sommeria ensemble requires specialized techniques \citep{AbMa:03,DuFr:10}, and much less is known about the accuracy of sampling the nonequilibrium steady states treated in this paper.

In the fluid dynamics setting, several equilibrium models are known.  For unforced, ideal fluids in two dimensions, the Miller-Robert-Sommeria measure \citep{Miller90,Robert91,RoSo91}, which encodes the area distribution of the vorticity field, is well established \citep{BoVe12}.  In fluctuating hydrodynamics, the Landau-Lifshitz-Navier-Stokes equations are provably ergodic with respect to the Boltzmann-Gibbs distribution of kinetic energy under a fluctuation-dissipation relation and stochastic forcing \citep{EMa:01,EmaSi01,DoVaGaBe10,DeGrVaDo13}. 
For the geophysically relevant regime of fixed-wavelength stationary forcing and viscous dissipation, fluids in the atmosphere and ocean are believed to sample a nonequilibrium steady state in which the kinetic energy spectrum satisfies a power law over a range of length scales, as posited by Kolmogorov \citep{Leith:68,Batchelor:69,Kraichnan:71,Frisch:95,TuSm06}.  In this case no equilibrium measure is explicitly known, but in the probabilistic setting a stationary \emph{expectation}---namely the power-law spectrum for kinetic energy---can be observed from measurements in the atmosphere \citep{Nastrom:84}.

Power-law kinetic energy spectra can be simulated using forced Navier-Stokes discretizations at operational resolutions, but this typically requires the introduction of a viscosity coefficient that far exceeds that encountered in atmosphere/ocean science applications in nature.  In practice, higher order hyperviscosity is used because it has a more localized effect on the spectrum.  Excessive numerical viscosity is believed to adversely affect the simulated growth of small-scale physical instabilities as well as inhibiting spread in ensemble simulations \citep{KeThWo:12,ThKeWo:14}.  Our simulation experiments bear this out, as we observe a strong influence of numerical viscosity on autocorrelation functions and the information content of ensembles. 
Turbulent backscatter methods have been introduced in \citep{DoSa:97,Shutts:05,Berner:09} to re-inject kinetic energy at viscous length scales.  Alternatively, ``superparameterization'' methods \citep{XiMaGr09,GrMa2013} have been proposed as an intermediate alternative to large eddy simulations.  In these, eddy dynamics are modelled by either a simplified dynamics or a stochastic closure model.
In this paper we adopt an extreme statistical simplification of the fine-scale model, coupling it via thermostatic controls to directly impose a background power law kinetic energy spectrum at the smallest resolved scales.   Our approach allows us to maintain the given target without employing artificially increased viscosity. The energy spectrum we impose can be taken from observational data, theory or higher resolution simulations. In the case of atmospheric turbulence the experiments by \citet{NaGa:85} provide such data. In two-dimensional forced-dissipated Navier-Stokes it may also be taken from theoretical predictions \citep{Leith:68,Batchelor:69,Kraichnan:71}.  

The remainder of this paper is organized as follows: In the following section the incompressible 2D Navier-Stokes equations, with forcing and dissipation, are recalled in their vorticity form. Section \ref{sec:canon} discusses the dynamical perturbations used in molecular dynamics to simulate a molecular gas at constant temperature. The interpretation of such perturbation methods in the context of turbulence is discussed in Section \ref{sec:invariant}. A feedback control is then applied to 2D turbulence simulations in Sections \ref{sec:c2d} and \ref{sec:cbs}; the former is a simulation with large scale random forcing and forward enstrophy cascade and the latter is a simulation augmented with additional, small scale forcing that is unresolved due to spectral truncation. Both of these sections include statistic and dynamic results of the new approach. A short discussion of similar methods and possible practical applications in Section \ref{sec:discussion} concludes the paper.

\section{Two-dimensional turbulence}
\label{sec:2dt}
We focus on driven two-dimensional incompressible flow.  Ignoring rotation and topographical effects, we work with the Navier-Stokes equations on a doubly periodic domain $\bs{x}\in\mathbb{T}^2$.  
The 2D Navier-Stokes equations with forcing $f(\bs{x},t)$ and generalized viscosity model are:
\begin{equation}
 \omega_t + J(\psi,\omega) = f + \sum_{j \in \mathbb{J} \subset \mathbb{Z}} \nu_j \Delta^j \omega,	\qquad \Delta \psi = \omega,
 \label{eq:2dNS}
\end{equation}
where $\Delta$ is the Laplace operator, $\psi(\bs{x},t)$ is the stream function and $\omega(\bs{x},t)$ is the vorticity component normal to the plane.  The nonlinearity $J(\psi,\omega)$ is defined by
\begin{equation}
  J(\psi,\omega) = \frac{\partial \psi}{\partial x}\frac{\partial \omega}{\partial y}  - 
  \frac{\partial \psi}{\partial y}\frac{\partial \omega}{\partial x}. \label{eq:J}
\end{equation}
The summation on the right side of the first relation in \eqref{eq:2dNS} is a general linear viscosity model.
In this paper we use two viscosity terms, $\mathbb{J}=\{-1,p\}$.  These include either physical viscosity, $p=1$ or fourth order hyperviscosity ($p=4$), and a hypoviscous term with $j=-1$ to curtail the inverse energy cascade. The latter term models bottom friction in atmospheric fluid flow \citep{DaGu:00}.

Equation \eqref{eq:2dNS} is discretized using a pseudo-spectral method \citep{Canuto:06}, expressing the vorticity field in terms of its Fourier components
\begin{equation}
 \omega_{\bs{k}} = \frac{1}{(2\upi)^2} \int_{\mathbb{T}^2} \omega(\bs{x}) e^{-\ui\bs{k}\cdot\bs{x}} \dd \bs{x},
 \quad -K \le k_1, k_2 \le K,
 \label{eq:Fourier}
\end{equation}
where $\bs{k} = (k_1, k_2)$,  is an index vector, and we denote $|\bs{k}| = (k_1^2+k_2^2)^{1/2}$ and $|\bs{k}|_\infty = \max \{ |k_1|, |k_2| \}$.
In terms of its Fourier components, equation \eqref{eq:2dNS} is written
\begin{equation}
 \dot{\omega}_{\bs{k}} + J_{\bs{k}}(\bs{\omega}) = f_{\bs{k}} + \nu_1 \Delta_{\bs{k}} \omega_{\bs{k}} + \nu_{-p} \Delta_{\bs{k}}^{-p} \omega_{\bs{k}},  \label{discrete}
\end{equation}
where $\Delta_{\bs{k}} = -|\bs{k}|^2$ and $J_{\bs{k}}(\bs{\omega})$ represents the pseudo-spectral evaluation of the nonlinear term \eqref{eq:J} on a uniform $2K \times 2K$ grid, implementing a standard 3/2 de-aliasing filter \citep{Canuto:06}. 


Our computational set-up is similar to that of \citet{Gotoh:98}.  The effect of the hypoviscosity is restricted to those modes with $|\bs{k}| \leq 3$.  The forcing is Gaussian white noise in time and applied in a band of energy shells with $3.5 < |\bs{k}| < 6.5$. For the simulations of Section \ref{sec:cbs}, small scale random forcing is additionally applied in the range $202.5 < |\bs{k}| < 206.5$.
The magnitude of the forcing is scaled such that the expected power input matches a given value $P$; see Appendix \ref{sec:forcing} for details. The values of the parameters used in the simulations are summarized in table \ref{tab:pars}.

\begin{table}
  \begin{center}
  \def~{\hphantom{0}}
  \begin{tabular*}{\textwidth}{@{\extracolsep{\fill} } lcccc}
Physical & Reference &  Truncated & Hyperviscosity & Nos\'e-Hoover \\
\hline
$P$ & $0.1$ & $0.1$ & $0.1$ & $0.1$ \\
$\nu_{-1}$ & $2$ & $2$ & $2$ & $2$ \\
$p$ & $1$ & $1$ & $4$ & $1$ \\
$\nu_p$ & $1.0 \times 10^{-4}$ & $1.0 \times 10^{-4}$ & $4.3 \times 10^{-15}$ & $1.0 \times 10^{-4}$ \\
\hline
Numerical & & & & \\
\hline
$\Delta t \times 10^3$ & $1$ & $1$ & $1$ & $1$ \\
$K$ & $256$ & $85$ & $85$ & $85$ \\
$\ell^*$ & - & - & - & $\lbrace 51,71^\dagger,81 \rbrace$\\
$\varepsilon_0$ & - & - & - & $\lbrace 10^{-1},10^{-1/2},1^\dagger \rbrace$ \\
\hline
Results & \multicolumn{4}{r}{(case denoted $\dagger$)} \\
\hline
$\eta$ & $4.92$ & $4.92$ & $4.92$ & $4.92$ \\
$C'$ & $1.15$ & $0.645$ & $1.33$ & $1.14$ \\
$d$ & $0.789$ & $0.305$ & $0.900$ & $0.779$ \\
  \end{tabular*}
  \caption{Parameters and results for the simulation of two-dimensional turbulence}
  \label{tab:pars}
  \end{center}
\end{table}

The viscous terms in (\ref{discrete}) have typical length scales defined by the wave number magnitudes for which the coefficients $\nu_p \Delta_{\bs{k}}^p$ have magnitude unity. 
By using $\Delta_{\bs{k}} = -|\bs{k}|^2$ we find $k_d = \sqrt{\nu_1}^{-1}$ for $j=1$ and $k_h = \sqrt{\nu_{-1}}$ for $j=-1$. Assume that these two scales are sufficiently well separated, $k_d \gg k_h$, and that forcing acts primarily at some intermediate length scale. In this setting it is expected that the hypothesis of \citet{Kraichnan:67} holds and that there is a steady flux of energy from the forcing wave numbers to larger scales (i.e.~the inverse energy cascade), as well as a steady enstrophy flux to smaller scales (i.e.~the direct enstrophy cascade). These cascades terminate when the dissipative scales $k_h$ and $k_d$ are reached, but with sufficient separation between forcing and dissipation scales there persist a range of wave numbers of statistically stationary energy and enstrophy transport. Because the dynamics in these ranges are almost unaffected by damping and forcing, they are dubbed inertial ranges. The steady fluxes of energy (resp.~enstrophy) in both regimes yield  power law energy spectra.
 
This means the energy in wavenumbers near $k$, given by 
\begin{equation}
 E_k(\bs{\omega}) =  -\half \sum_{k-\half < |\bs{k}| < k+\half} \Delta_{\bs{k}}^{-1} \omega_{\bs{k}} \omega_{\bs{k}}^*, \label{eq:kes}
\end{equation}
satisfies on average an approximate power law (time average denoted by overbar) 
\begin{equation}
  \bar{E}_k \approx C \epsilon^{2/3} k^{-5/3} \mbox{ for } k \ll k_f,
\end{equation}
in the inverse energy cascade, and
\begin{equation}
  \bar{E}_k \approx C \eta^{2/3} k^{-3} \mbox{ for } k \gg k_f,
\end{equation}
in the direct enstrophy cascade. The parameters $\epsilon$ and $\eta \approx k_f^2 \epsilon$ denote the energy and enstrophy injection rates \citep{Bofetta:10}.
Whether or not the kinetic energy spectrum observed in the atmosphere can be explained by such two-dimensional turbulence assumptions is debated \citep{Lindborg:99}. 
The methodology we propose makes no assumptions on the functional form of the kinetic energy spectrum. Therefore it is applicable to any observed spectrum. We demonstrate this by using it in a forced-dissipated turbulence cascade in Section \ref{sec:c2d} and in a case with both large and small-scale stochastic forcing in Section \ref{sec:cbs}. The versatility of the method also promises straightforward generalization to three-dimensional turbulence.

\begin{figure}
  \begin{center}
    \includegraphics[width=.6\textwidth]{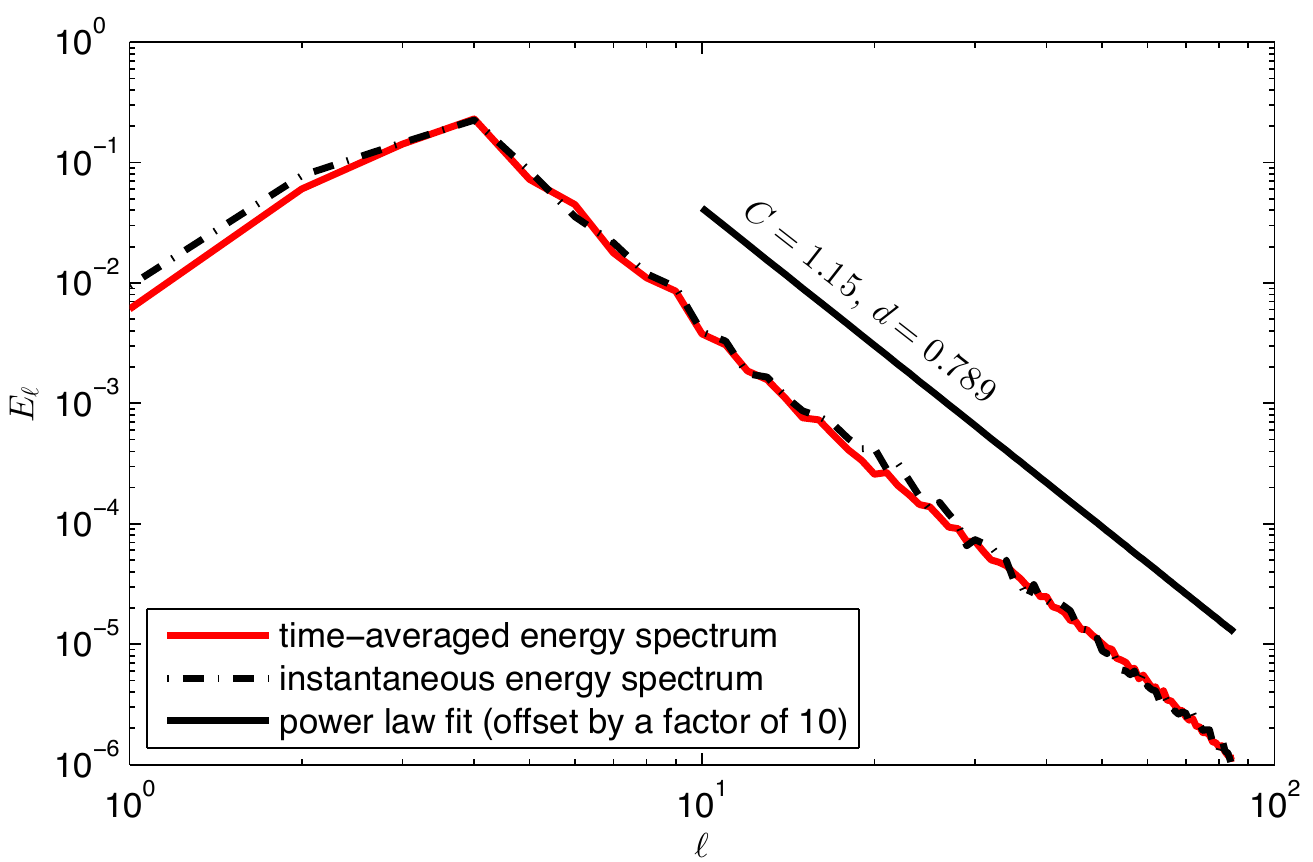}
  \end{center}%
  \caption{Two-dimensional turbulence kinetic energy spectrum and power law fit. Simulation parameters are given in the ``reference'' column of table \ref{tab:pars}.}
  \label{fig:turb_spect}
\end{figure}

The effect of spectral truncation on the kinetic energy spectrum is most pronounced in the inertial enstrophy regime. To save computational effort, we design our simulation with small separation between the forcing scale and the hypodiffusion scale. Such a parameter set is given in the column labeled ``Reference'' of table \ref{tab:pars}. This simulation yields an inertial range power spectrum as depicted in figure \ref{fig:turb_spect}.  The figure shows both the instantaneous spectrum and the time-averaged spectrum after $50$ time units, corresponding to over 250 eddy turnovers. 
The computed energy spectrum is steeper than the hypothesized $k^{-3}$ slope. This is common in numerical simulations and is usually attributed to insufficiently large Reynolds number due to limited resolution \citep{Bofetta:10,Gotoh:98,BrMcMuPrWe:00}. \citet{Saffman:71} suggests a $k^{-4}$ spectrum as the result of small discontinuities at the smallest scales. \citet{FaKePr:11} suggest that numerical simulations differ from the hypothesized spectra not due to insufficient resolution, but due to the choice of forcing.  They investigate forcing functions that yield the hypothesized $k^{-5/3}$ and  $k^{-3}$ power law regimes.
This is in line with the findings of \citet{DaGu:00}, who demonstrate that power law spectra can only be observed for a special set of external parameters.  \citet{FaKePeGo:96} use wavelet methods to analyse turbulent velocity fields and define a local energy spectrum. They discover that the $k^{-3}$ energy spectrum only holds outside of regions of strong vorticity and shear layers. Inside those regions the energy spectrum scales as $k^{-6}$ and $k^{-4}$ respectively. \citet{SuGaCh:99} find that the forward enstrophy cascade spectral slope depends directly on the chosen large scale drag.

We fit a power law to the observed (steeper) spectrum in the form suggested by \citet{Gotoh:98}
\begin{equation}
  \bar{E}_k \approx C' \eta^{2/3} k^{-3-d},
\end{equation}
using a least-squares approach. The newly introduced parameter $d$ indicates a deviation from the theoretical slope. Even though the power law that develops for a given parameter set in \eqref{eq:2dNS} is different from the theoretical spectrum, it is still independent of the chosen initial conditions. This is evidence that the dynamics are ergodic and motivates consideration of the invariant measure of the dynamical system, analogous to such considerations in molecular dynamics. The following section briefly consider the tools used in molecular dynamics, before their application to 2D turbulence is detailed in Section \ref{sec:invariant}.

\section{Canonical sampling and temperature control}
\label{sec:canon}
In this section we briefly recall the problem of constant temperature simulations in molecular dynamics, which is the inspiration for the control we propose for kinetic energy in the Navier-Stokes equations. 
The dynamics of a classical molecular gas are governed by a Hamiltonian system
\begin{align}
  \dot{\bs{q}} &= \bs{p}, \label{md1}\\
  \dot{\bs{p}} &= -\nabla_{\bs{q}} V(\bs{q}), \label{md2}
\end{align}
where $\bs{q} \in \mathbb{R}^n$ represents the vector of particle positions, $\bs{p} \in \mathbb{R}^n$ the vector of particle momenta, and unit mass is assumed.  The Hamiltonian $H(\bs{q},\bs{p}) = \frac{1}{2} \| \bs{p} \|^2 + V(q)$ represents the total energy as a sum of kinetic energy $\kappa(\bs{p})=\frac{1}{2} \| \bs{p}\|^2$ and potential energy $V(\bs{q})$.  
Given an initial condition with total energy $H_0$, a solution is restricted to the constant energy surface $H(\bs{q}(t),\bs{p}(t))=H_0$.  If the phase flow is ergodic on the constant energy surface, then for almost any initial condition, the time average of an observable function $a(\bs{q},\bs{p})$ of the solution,
\begin{equation}
  \bar{a}(\bs{q},\bs{p}) = \lim_{T\rightarrow\infty} \frac{1}{T}\int_0^T a(\bs{q}(t),\bs{p}(t)) \dd t,
\end{equation}
is equal to the ensemble average with respect to the \emph{microcanonical} measure $\pi_m$,
\[
	\langle a \rangle = \int a(\bs{q},\bs{p}) \, \pi_m (\dd q,\dd p ) = \int a(\bs{q},\bs{p}) \, \delta (H(\bs{q},\bs{p})-H_0) \, \dd \bs{q} \, \dd \bs{p},
\] 
where $H_0$ is the total energy defined by the initial condition.  

For a molecular gas in thermal equilibrium with a reservoir of temperature $\tau$, the total energy is no longer conserved.  Instead the time averaged kinetic energy satisfies
\begin{equation}
	 \bar{\kappa}= \frac{n k_B \tau}{2}= \frac{n}{2\beta},
\end{equation}
where $k_B$ is the Boltzmann constant and $\beta$ is the inverse temperature.
In this case, trajectories of the system sample the \emph{canonical} (Gibbs) measure
\[
	\pi_c(\dd \bs{q},\dd \bs{p}) = \rho_c(\bs{q},\bs{p}) \, \dd \bs{q}\,\dd \bs{p}, \quad \rho_c(\bs{q},\bs{p}) \propto e^{-\beta H(\bs{q},\bs{p})},
\]
and when the flow is ergodic in this measure the temperature is related to the canonical mean
\[
	\frac{n k_B \tau}{2} = \langle \kappa \rangle = \int \kappa(\bs{p}) \pi_c (\dd \bs{q} , \dd \bs{p}).
\]

To carry out numerical simulations of molecular dynamics at constant energy, there exist numerical methods that (exactly or approximately) preserve the Hamiltonian.  Preservation of a quantity like $\bar{\kappa}$ is more subtle, since it is only conserved `on average'.  Methods for constant temperature molecular dynamics introduce perturbations to the dynamical equations \eqref{md1}--\eqref{md2} called thermostats.  
The Nos\'e-Hoover thermostat \citep{Nose84a,Nose84b,Hoover85}  augments the dynamics with an extra variable $\xi$ that controls the kinetic energy in the system as follows
\begin{align}
  \dot{\bs{q}} &= \bs{p}  \label{eq:nh1} \\
  \dot{\bs{p}} &= -\nabla_{\bs{q}} V(\bs{q})  - \xi \varepsilon \bs{p}  \label{eq:nh2} \\
  \dot{\xi}    &= \varepsilon \left( \beta \kappa(\bs{p}) - n\right),  \label{eq:nh3} 
\end{align}
where $\varepsilon$ is a coupling parameter.  The Nos\'e-Hoover method can be interpreted in two ways.  First, it is straightforward to verify that the Liouville equation associated to the augmented system \eqref{eq:nh1}--\eqref{eq:nh3} admits the steady state
\begin{equation}
  \rho(\bs{q},\bs{p},\xi) = \rho_c(\bs{q},\bs{p}) e^{-\half \xi^2},
\end{equation}
for which the marginal density with respect to $\bs{q}$ and $\bs{p}  $ is the Gibbs distribution.  Hence the Nos\'e-Hoover method enforces the canonical invariant measure.  On the other hand, it is also apparent that the variable $\xi$ acts as a damping coefficient for kinetic energy when $\xi>0$ and excites kinetic energy for $\xi<0$.  Furthermore, \eqref{eq:nh3} shows that $\xi$ will increase (decrease) when $\kappa(\bs{p})/n$ exceeds (falls short of) the target temperature $\beta^{-1} = k_B \tau$.  Hence, the Nos\'e-Hoover method can also be interpreted as a \emph{negative feedback control} on temperature. This second interpretation is crucial to our application of the thermostat to fluids, as we explain below.

\section{Invariant measures and expectations for two-dimensional turbulence}
\label{sec:invariant}
The tendency of forced-dissipated turbulence to develop a power law spectrum independent of initial conditions provides evidence that the dynamics may sample a unique invariant measure. Ergodicity of finite truncations of the forced-dissipated two-dimensional Navier-Stokes equations is proven by \citet{EMa:01} for the case of stochastic forcing of only two long wave modes, $\bs{k}_1=(1,1)$ and either $\bs{k}_2=(1,0)$ or $\bs{k}_2=(0,1)$. Continuing this analysis, it is readily proven that this result extends to the case of forcing arbitrary wave numbers $\bs{k}_1$ and $\bs{k}_2$ as well as $\bs{k}_1 + (1,0)$ and $\bs{k}_2 + (0,1)$. This is the case in our proposed forcing of selected bands in Fourier space.

Our interest lies in the practical case where computational costs prohibit resolving a sufficient number of modes to capture small-scale dissipation. This is the case in many large scale atmosphere and ocean applications. The computational load is determined by restricting the Fourier expansion in \eqref{eq:Fourier} to those modes with $|\bs{k}|_{\infty} \leq K$. Denote the truncated vector of vorticity coefficients by $\bs{\omega}_{K}$, and the full Fourier transform by $\bs{\omega}_{\infty}$. If the resolution $K$ is insufficiently large to capture the diffusion of enstrophy at the scale $k_d$, the dynamics of the truncated system will differ greatly from those in a system that is well-resolved. 

Stochastic approaches to model the effect of the unresolved degrees of freedom focus on the dynamical interaction between resolved and unresolved modes (see for instance
\citet{Mori65,Zwanzig61,Hasselman76,MaTiVa01,FaVa04}).
Here, we instead focus on correcting the statistics of the truncated system, as embodied in its invariant distribution and expectations.

Ideally, for correct sampling we would require that the invariant density $\rho_{K}$ sampled by the truncated dynamics (i.e.,~$\bs{\omega}_K \sim \rho_{K}$) be equal to the marginal distribution of those same modes in the resolved case. Let us introduce a partition $\bs{\omega}_{\infty} = (\hat{\bs{\omega}},\tilde{\bs{\omega}})$ where $\hat{\bs{\omega}}$ consists of resolved modes with $|\bs{k}|_\infty < K$ and $\tilde{\bs{\omega}}$ for the unresolved modes. For equivalence between the invariant measures of the two systems, we would require $\rho_K$ to be equal to the marginal density
\begin{equation}
	\hat\rho = \int \rho_\infty(\hat{\bs{\omega}},\tilde{\bs{\omega}}) \dd \tilde{\bs{\omega}}.
\end{equation} 
However, steady solutions to the Fokker-Planck equation for general forced-dissipated turbulence are not explicitly available, due to the inherent complexity of the nonlinear wave interactions. This means we cannot define a perturbation for the truncated system such that the invariant measure is correct.

Instead we focus attention on \emph{expectations} in the unknown measures.  If the invariant measures $\bar\rho$ and $\rho_K$ \emph{did} match, so would the expectations of arbitrary observables $\ensbra{a(\bar\omega)}$. Ergodicity of the systems would then also imply equivalent time averages.
We take the energy spectrum as a set of observables. Given a truncation $K\gg k_d$, the modes $\omega_{\bs{k}}$ with $k_f \ll |\bs{k}| \ll k_d$ revert to a power law spectrum due to the forward enstrophy cascade. In a system truncated well below the viscous scale $K\ll k_d$, the downscale cascade of enstrophy is terminated abruptly, resulting in an artificial build-up of enstrophy at the smallest resolved scales known as spectral blocking. 
An inaccurate energy spectrum in the highest wave numbers eventually leads to deviation from the power law spectrum in the energy range, i.e. in the large scales \citep{SuGaCh:99}.

In this paper, we propose employing the Nos\'{e}-Hoover thermostat (\ref{eq:nh1})--(\ref{eq:nh3}) to enforce a power law spectrum on the kinetic energy in the absence of a mean flow. 
A crucial difference in the application to fluids, compared to constant temperature molecular dynamics, is that in the current context the invariant measure of the extended variable is unknown.  In canonical sampling,  the distribution for the thermostat variable $\xi$ is known to be normally distributed with mean zero, independent of the original dynamics. This allows augmentation of the thermostat variable dynamics \eqref{eq:nh3} by an Ornstein-Uhlenbeck process without disrupting the invariant measure \citep{SaDeCh07}. The addition of stochastic forcing in the auxiliary variable makes the thermostat ergodic  \citep{Leimkuhler:09}. However, in the current context of forced turbulence, we do not know the distribution of the thermostat variable. Moreover, it is expected that $\xi(t)$ will have nonzero mean, because it must remove excess energy on average, yet add energy in the form of backscatter on occasion. For this reason it is crucial to \emph{exclude} the stochastic process in the thermostated wave numbers. 
Similar arguments were used in a forced molecular model in \citet{JoLe11}.

\section{Feedback control of the forward enstrophy cascade}
\label{sec:c2d}
The kinetic energy spectrum consists of the kinetic energy distributed over energy shells in wavenumber space, see \eqref{eq:kes}. A Nos\'e-Hoover thermostat could be applied to each shell to drive its energy to the observed average. However, it is undesirable to artificially perturb the largest scale modes in the system, which are well-resolved and least uncertain. For this reason, only energy shells with wave number $\ell>\ell^*$ above a threshold are equipped with a Nos\'e-Hoover thermostat.

The discrete equations of motion for two-dimensional Euler flow, extended with the thermostats, reads
\begin{align}
 \dot{\omega}_{\bs{k}} + J_{\bs{k}}(\bs{\omega}) &= f_{\bs{k}} + \nu_1 \Delta_{\bs{k}} \omega_{\bs{k}} + \nu_{-1} \Delta_{\bs{k}}^{-1} \omega_{\bs{k}} - \sum_{\ell=\ell^*}^{\ell_\text{max}} \xi_\ell \varepsilon_\ell \partial_{\bs{k}} E_\ell(\bs{\omega}) \\
\dot{\xi}_\ell &= \varepsilon_\ell \left( E_\ell(\bs{\omega}) - \bar{E}_\ell \right), \quad \ell = \ell^*,\ldots,\ell_\text{max}. \label{eq:kNH}
\end{align}
The perturbation parameters $\varepsilon_\ell$ are chosen such that $\varepsilon_\ell^{2} / \bar{E}_\ell = \varepsilon_0^{2}$ for all $\ell$.  After substituting the waveband energy we may write
\begin{align}
 \dot{\omega}_{\bs{k}} + J_{\bs{k}}(\bs{\omega}) &= f_{\bs{k}} + \nu_1 \Delta_{\bs{k}} \omega_{\bs{k}} + \nu_{-1} \Delta_{\bs{k}}^{-1} \omega_{\bs{k}} + \sum_{\ell=\ell^*}^{\ell_\text{max}} \xi_\ell \varepsilon_\ell \sum_{\ell-\half < |\bs{k}| < \ell+\half} \Delta_{\bs{k}}^{-1} \omega_{\bs{k}} \\
\dot{\xi}_\ell &= -\varepsilon_\ell \left( \half \sum_{\ell-\half < |\bs{k}| < \ell+\half} \Delta_{\bs{k}}^{-1} \omega_{\bs{k}} \omega_{\bs{k}}^* - \bar{E}_\ell\right), \quad \ell = \ell^*,\ldots,\ell_\text{max}. \label{eq:kNH_subs}
\end{align}
This form makes clear that each $\xi_\ell$ acts as a dissipation/growth coefficient. 

For each thermostatted mode, the energy is driven towards a target value. We emphasize that this value may be taken from {physical observations, theoretical predictions or, as here for the purpose of method evaluation, from a high fidelity solution that resolves the physical viscosity. As such the method may be seen as a data assimilation approach that uses statistical data to correct mean statistics of a dynamical simulation.

It should be noted that while the control will certainly drive the system towards correct averages for the energy levels in the thermostatted energy shells, the invariant measure sampled by the trajectories remains unknown. An important consequence of this is that the marginal distribution of the thermostat variables is not known a priori. This complicates choosing initial values for the thermostat variables, as initialising them far from their equilibrium will result in a slow relaxation. We perform a pilot simulation in which the thermostatted system is allowed to fully equilibrate in order to select initial data.



\subsection{Energy spectrum}
The energy spectrum for a simulation using a thermostat is compared to a resolved model and an underresolved model with hyperviscosity ($\nu_4 = 4.3\times10^{-15}$) in figure \ref{fig:spectrum_NH}. 
The kinetic energy per energy shell is multiplied by a correction factor that accounts for (i) the nonuniformity that arises by partitioning of the discrete Fourier space into annular shells and (ii) the incomplete resolution of the highest wave number bands (i.e.~those in the corners of the Fourier space). The factor is $\frac{\upi}{\ell W_\ell}$, where $W_\ell = -\half \sum_{\ell-\half < |\bs{k}| < \ell+\half} \Delta_{\bs{k}}^{-1}$, and essentially results in smooth spectra in figure \ref{fig:spectrum_NH} (cf.~Fig.~\ref{fig:turb_spect}). The artifical (hyper)viscosity model grossly underestimates the kinetic energy in the large wave numbers compared to the well-resolved model. The kinetic energy spectrum in the thermostated model is visually indiscernible from the reference model. 
\begin{figure}
  \begin{center}
    \includegraphics[width=0.6\textwidth]{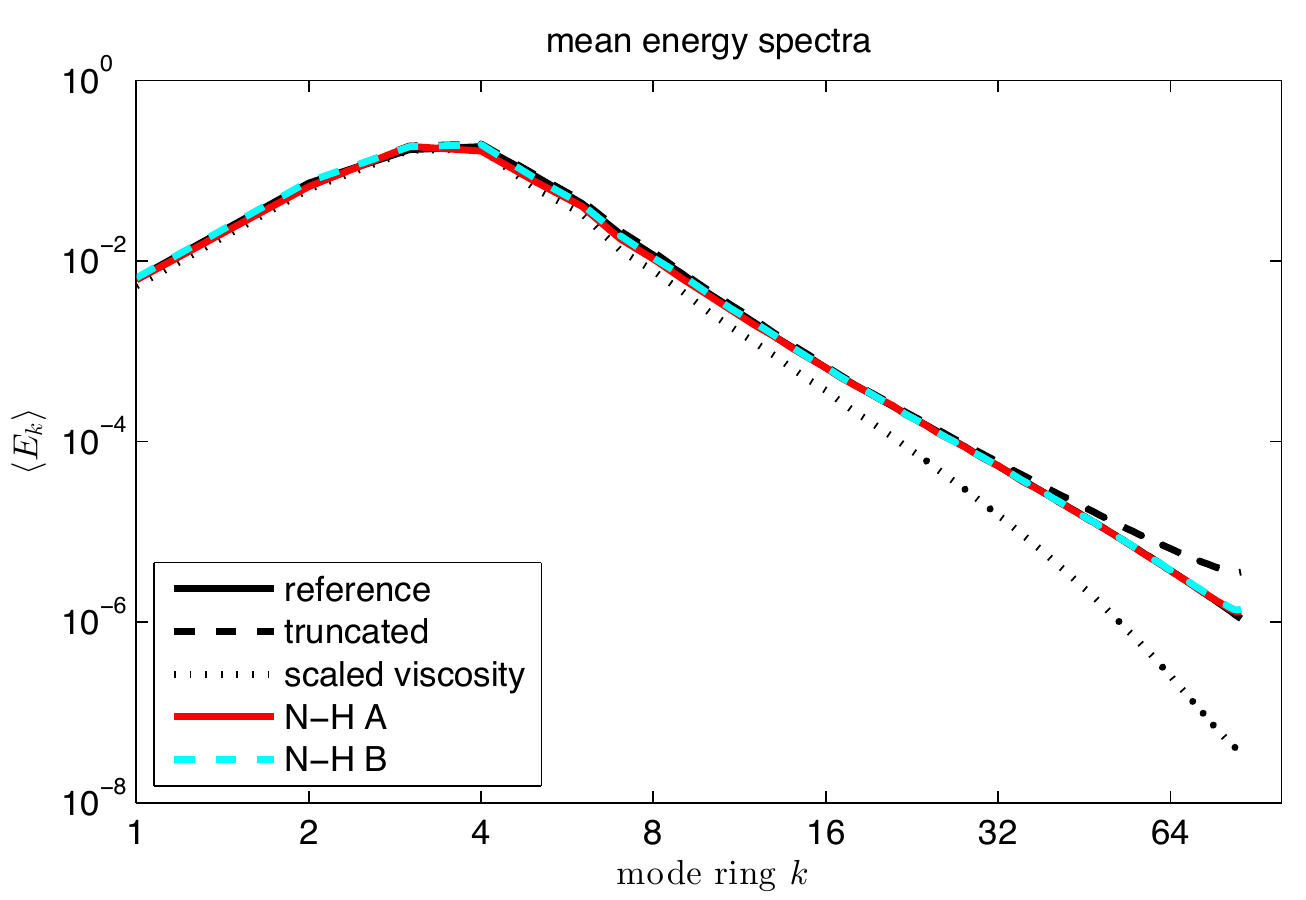}
  \end{center}%
  \caption{The kinetic energy spectrum after 100 time units for the reference simulation (solid), a truncated simulation (dashed), a simulation with hyperviscosity (dotted) and a simulation using the proposed control on the energy spectrum (dashed, red). Parameters are given in table \ref{tab:pars}, where the $\dagger$ indicates the parameters used here for the Nos\'e-Hoover control.}
  \label{fig:spectrum_NH}
\end{figure}

To investigate the energy spectrum more closely, 
the mean energy per mode is represented in figure \ref{fig:meanabsQ} for the reference, hyperviscosity, and thermostatted simulations. The mean energy per mode is multiplied by the cube of the wavenumber magnitude. This will result in areas of equal colour for modes where the energy spectrum scales as $k^{-4}$ and where isotropy can be assumed. Anisotropy would appear as a break of the radial symmetry. 
This does not occur in the reference simulation and for most of the controlled simulation. However, in figure \ref{fig:meanabsQ} we do observe that, for the thermostated model, kinetic energy appears to be more uniformly distributed across the Fourier modes within a given energy shell than is the case for the reference model. This is most likely due to the heavily reduced dimensionality of the phase space ($85^2$ versus $256^2$) leading to a much faster spread of the stochastic noise from the forcing through the available degrees of freedom.

\begin{figure}
  \begin{center}
    \includegraphics[width=\textwidth]{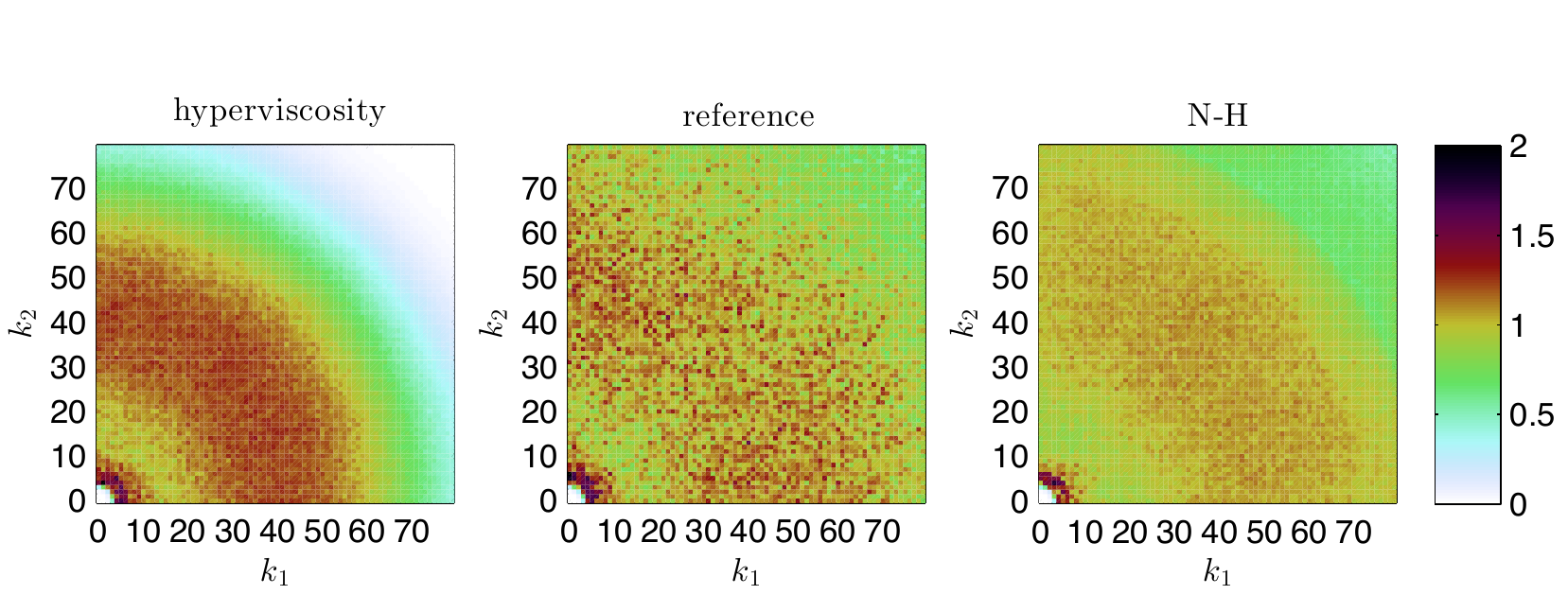}
  \end{center}
  \caption{A representation of the average energy in each Fourier mode using (left) the model with artificial viscosity, (centre) the reference simulation, and (right) the Nos\'e-Hoover simulation, see table \ref{tab:pars} for details. The value plotted is given by $\hat{C}_{\bs{k}} = \overline{\omega_{\bs{k}} \omega_{\bs{k}}^*} \frac{2 \upi}{C' \eta^{2/3}} k^{3+d+1}$, where the parameters $C'$ and $d$ are taken from the reference solution. When this value is close to unity, it indicates close local (in Fourier space) agreement to the power law spectrum.}
  \label{fig:meanabsQ}
\end{figure}

\subsection{Vorticity field}
In figure \ref{fig:vort_fields} the computed vorticity fields at $t=1$ and $t=10$ are shown for each of (centre) the reference model, (left) the model with artificial viscosity, and (right) the model using the Nos\'{e}-Hoover control. Both the hyperviscosity model and the control remain initially close to the reference solution. After 10 time units, the solutions have diverged due to the chaotic nature of the dynamics, but remain qualitatively similar.
\begin{figure}
  \begin{center}
    \includegraphics[width=\textwidth]{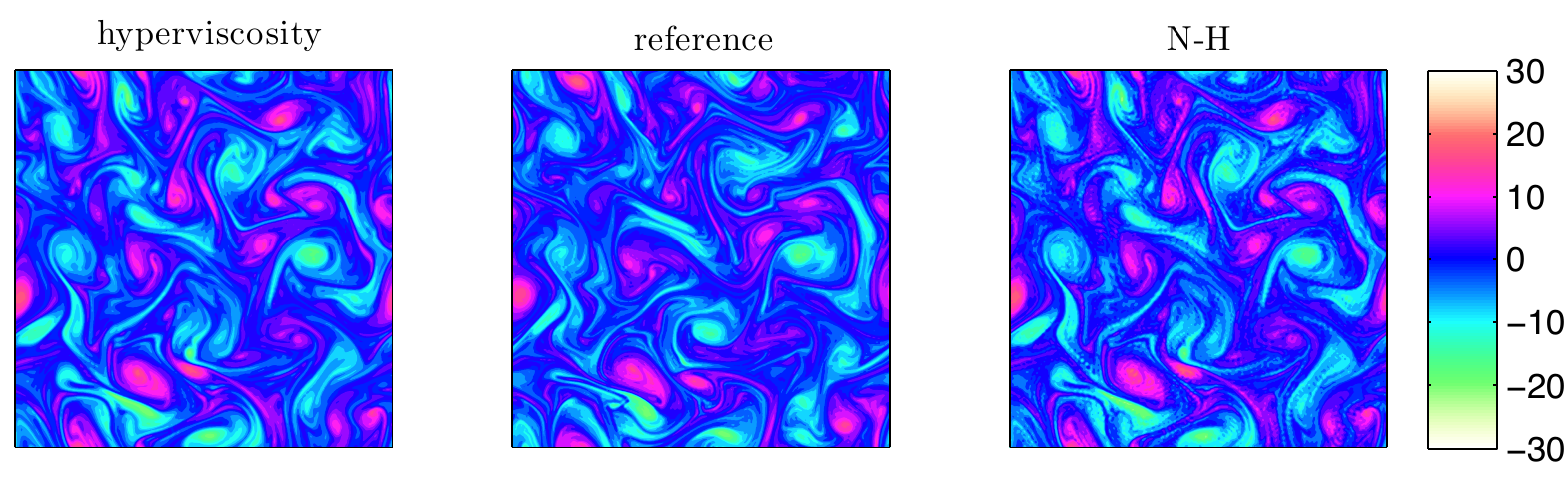}
    \includegraphics[width=\textwidth]{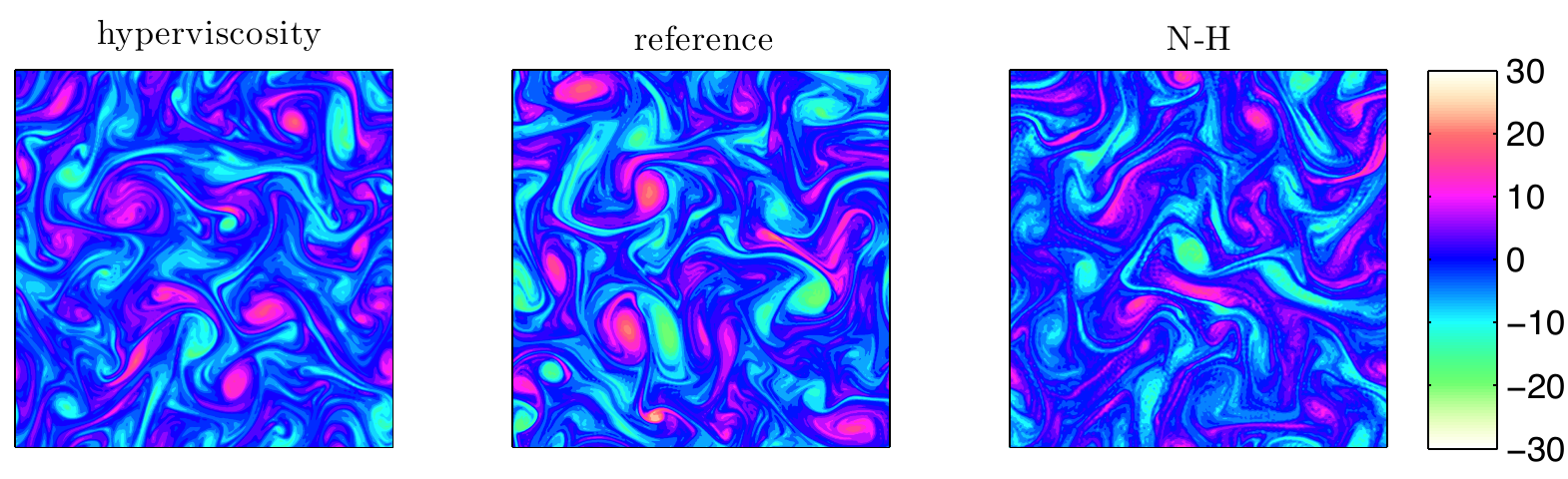}
  \end{center}%
  \caption{Vorticity fields in a classical double cascade, obtained at $t=1$ (top) and $t=10$ (bottom) using (left) the model with artificial viscosity added, (centre) the reference simulation, and (right) the Nos\'{e}-Hoover method as indicated with $\dagger$ in table \ref{tab:pars}.}
  \label{fig:vort_fields}
\end{figure}

\subsection{Autocorrelation functions}
Autocorrelation functions provide dynamical information on the temporal variance at different scales. The autocorrelation function $R_{\omega\omega}(s;x)$ of the vorticity  $\omega$ at a point $x$ is given by
\begin{equation}
  R_{\omega\omega}(s; x) = \frac{1}{T} \int_0^T \omega(t+s,x)\omega(t,x)\, \dd t
\end{equation}
when observed over time $T$. We study the autocorrelation function of the vorticity at a grid point in the vorticity field. To save computational effort we will focus on the vorticity field corresponding to Fourier modes with $|\bs{k}|_{\infty} < 16$. As the vorticity at each grid point is identically distributed assuming homogeneous forcing, we average the autocorrelation functions over space to speed up convergence.

We compare the autocorrelation functions for the thermostatted simulations to the fully resolved simulation, a truncated simulation, and a hyperviscosity model.
In the left panel of figure \ref{fig:acf_NH_mus} the autocorrelation function is shown for different perturbation parameters $\varepsilon_0$. The right panel focuses on the short-time behaviour.
The agreement is relatively insensitive to perturbation parameter for the range of values shown (an order of magnitude).  
For larger values of $\varepsilon_0$, the thermostat acts more strongly, approaching Langevin dynamics in the limit of large $\varepsilon_0$ \citep{FrGo11}.  For smaller 
values of $\varepsilon_0$ the thermostat becomes very weak, meaning the relaxation of the spectrum requires averaging on long times. The choice of thermostat threshold $\ell^*$ has an even smaller effect on the autocorrelation functions and is therefore not shown.
\begin{figure}
  \begin{center}
    \includegraphics[width=.48\textwidth]{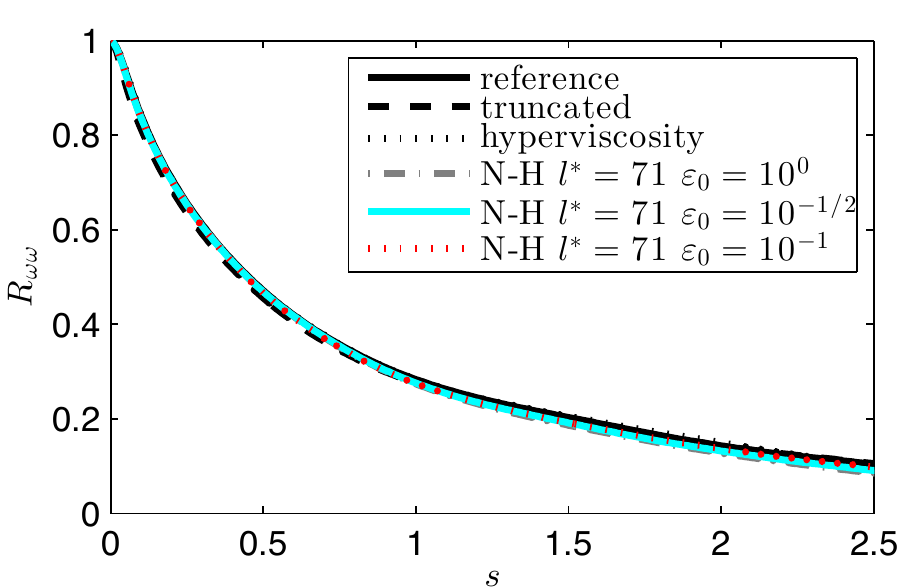}
    \includegraphics[width=.48\textwidth]{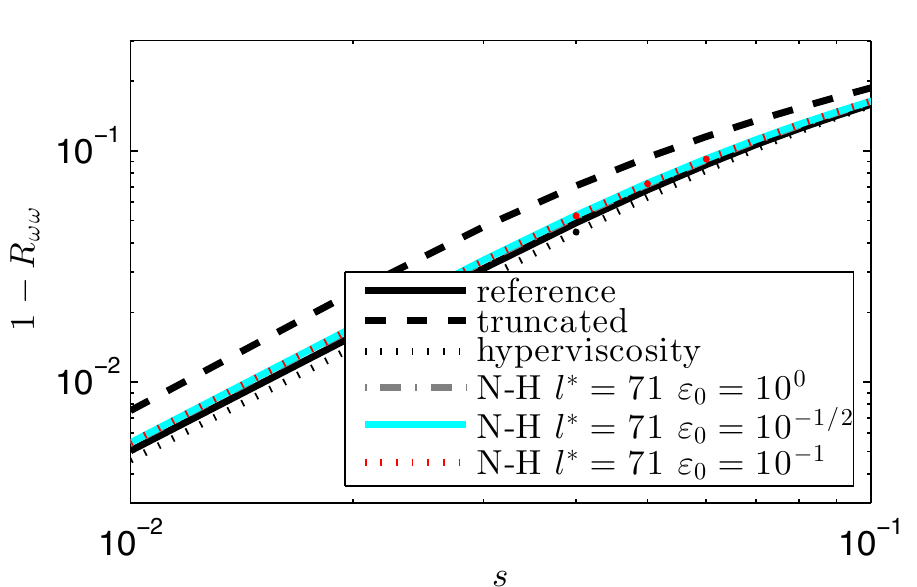}
  \end{center}%
  \caption{Autocorrelation functions $R_{\omega\omega}$ when simulating a classical double cascade. The Nos\'{e}-Hoover thermostat using various choices of the perturbation parameter $\varepsilon_0$ is compared to the reference solution (solid black), a truncated simulation (dashed black) and a hyperviscous model (dotted black).  The figure on the right shows $1-R_{\omega\omega}$ on a double logarithmic plot, to focus on short time scale behaviour.}
  \label{fig:acf_NH_mus}
\end{figure}

\subsection{Ensemble dispersion}
A 50-member ensemble is created from a single deterministic initial condition by randomizing the phase of all modes with $|\bs{k}| \geq 50$; in this way each ensemble member has an identical initial kinetic energy spectrum. 
In figure \ref{fig:ensemble} we compare ensembles, simulated up to $t=10$, for the reference simulation (centre), hyperviscosity (left) and Nos\'e-Hoover control (right) by studying the phase angle of the (0,1)-mode.
Both the hyperviscosity model and the Nos\'e-Hoover control are less dispersive than the reference solution. Nevertheless, the thermostatted ensemble exhibits observably more variance than the hyperviscosity model, and does manage to reflect some of the outlying trajectories of the reference solution.  

\begin{figure}
  \begin{center}
    \includegraphics[width=\textwidth]{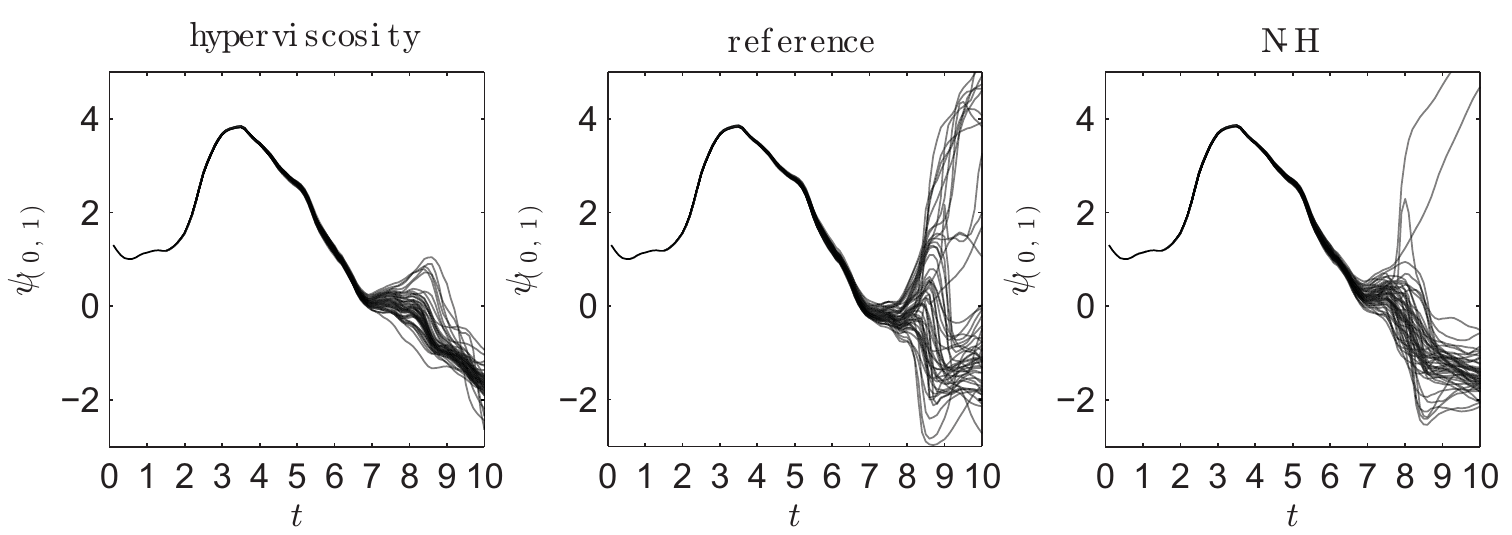}
  \end{center}%
  \caption{Ensemble dispersion in a classical double cascade. Plotted is the phase angle of the (0,1)-mode using  (left) the model with artificial viscosity, (centre) the reference simulation, and (right) the Nos\'{e}-Hoover method as indicated with $\dagger$ in table \ref{tab:pars}.}
  \label{fig:ensemble}
\end{figure}

\section{Feedback control of a system with subgrid scale forcing}
\label{sec:cbs}
In the previous section the Nos\'e-Hoover method corrected the energy spectrum in the forward enstrophy region for a truncated system. In this section we deviate from the classical setting of an intermediate forcing that results in two inertial ranges. Here we include a small-scale forcing term. This flattens the energy spectrum in the region between the two forcing scales when compared to the previous case. 
The form of the forcing is the same as before (Gaussian white noise in the Fourier components), only now the wave numbers with $202.5 < |\bs{k}| < 206.5$ are additionally forced. The total power input in these modes is equal to that of the low wave number forcing.

A feedback control with  remaining parameters as described in table \ref{tab:pars} is applied to a truncated simulation with $K=85$, i.e.~the small-scale forcing is \emph{not resolved}. Instead, the control target spectrum is observed from the fully resolved simulation after 100 time units. This simulates a scenario in which the fluid is forced at unresolved small scales, and we must attempt to incorporate this forcing given observations at resolved scales.
figure \ref{fig:spectrum_NH_2s} shows the mean kinetic energy spectrum after a 100 units for each of the four different simulations of table \ref{tab:pars}. Both the truncated and hyperviscosity models are oblivious to the small-scale forcing and consequently underestimate the energy in the smallest resolved scales. The Nos\'e-Hoover control acts only on mode bands $\ell^*$ and beyond ($\ell^* = 71$ in the figure), but still the energy spectrum is accurate over \emph{all} wave numbers.
\begin{figure}
  \begin{center}
    \includegraphics[width=0.6\textwidth]{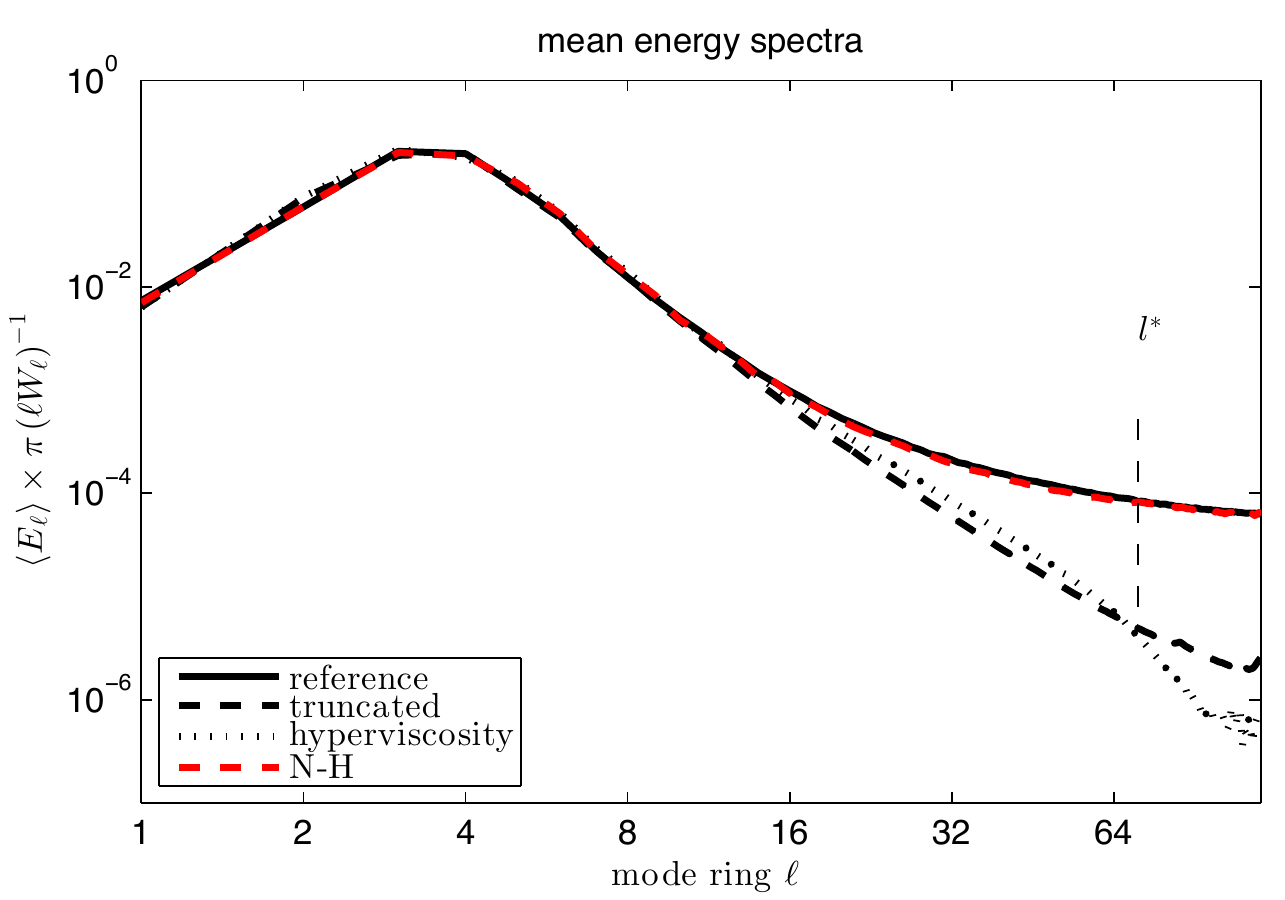}
  \end{center}%
  \caption{The kinetic energy spectrum after 100 time units for the reference simulation (solid), a truncated simulation (dashed), a simulation with hyperviscosity (dotted) and a simulation using the proposed control on the energy spectrum (dashed, red). Parameters are given in table \ref{tab:pars}, where the $\dagger$ indicates the parameters used here for the Nos\'e-Hoover control. An additional forcing is applied to the wavenumber with $203.5 < |\bs{k}| < 206.5$, with a power injection equal to that of the large scale forcing.}
  \label{fig:spectrum_NH_2s}
\end{figure}

\subsection{Vorticity field}
The inclusion of small-scale forcing leads to a noisier vorticity field for the reference solution as seen in the centre panels of figure \ref{fig:vort_fields_2s}. The hyperviscous vorticity field (left panels) is unable to capture this noise altogether. Using the Nos\'{e}-Hoover control produces similar vorticity fields. The chaotic nature of the dynamics leads to decorrelation of solutions over long time, yet at $t=10$ the controlled vorticity field is qualitatively still very similar to the reference.
\begin{figure}
  \begin{center}
    \includegraphics[width=\textwidth]{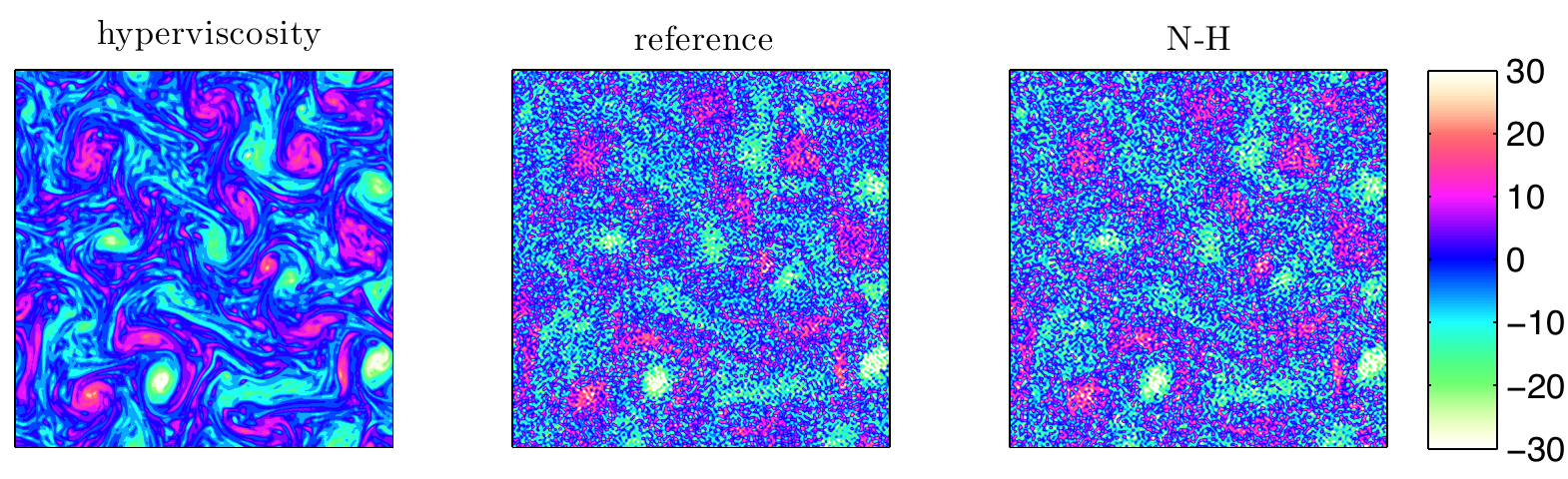}
    \includegraphics[width=\textwidth]{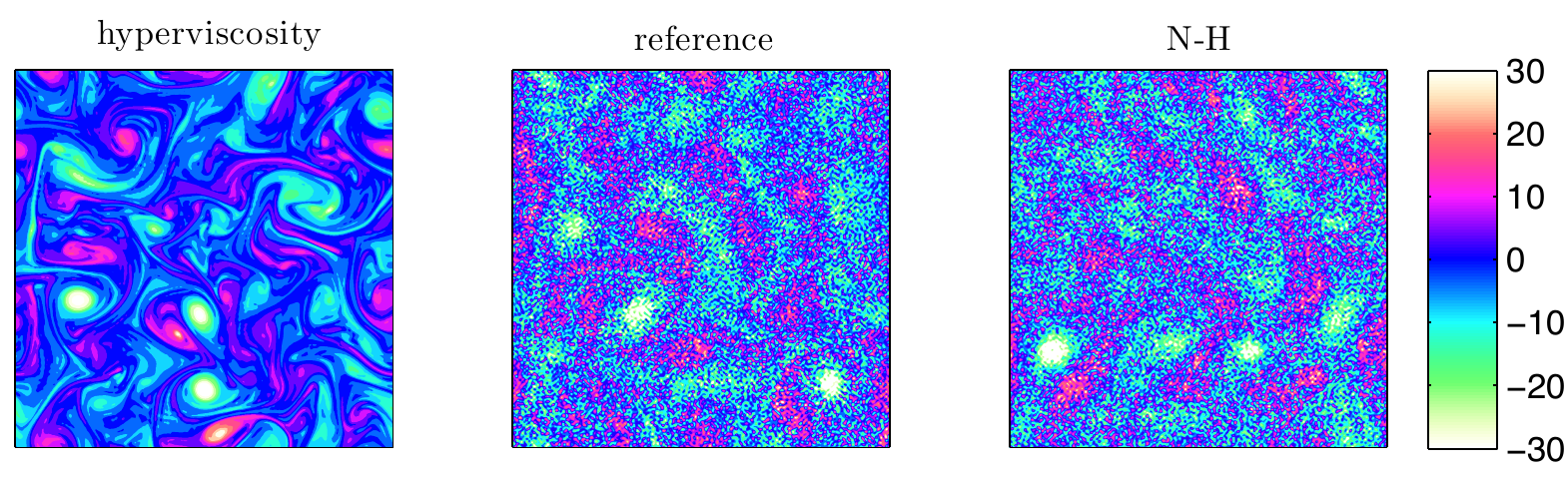}
  \end{center}%
  \caption{Vorticity fields with small scale forcing, obtained at $t=1$ (top) and $t=10$ (bottom) using (left) the overly diffusive hyperviscosity model, (centre) the reference simulation, and (right) the Nos\'{e}-Hoover multi-thermostatted method as indicated with $\dagger$ in table \ref{tab:pars}.}
  \label{fig:vort_fields_2s}
\end{figure}

\subsection{Autocorrelation functions}
As in the case with solely large scale forcing, we use autocorrelation functions for comparing dynamical properties.
In figure \ref{fig:acf_NH_mus_2s} we compare the Nos\'e-Hoover control with different perturbation parameters ($\varepsilon_0 \in \left\lbrace 1, 10^{-1/2}, 10^{-1} \right\rbrace$) against a reference simulation, a truncated simulation and a hyperviscosity model.
The results for the truncated and hyperviscous models show excessive correlation in time. The autocorrelation function for the controlled dynamics depends strongly on the perturbation parameter $\varepsilon_0$ in this case with small scale forcing. For smaller $\varepsilon_0$ the autocorrelation functions approach those of the truncated dynamics. The largest $\varepsilon_0$ considered does decorrelate similarly to the reference solution. The results are insensitive to the wavenumber threshold $\ell^*$ for the control.
\begin{figure}
  \begin{center}
    \includegraphics[width=.48\textwidth]{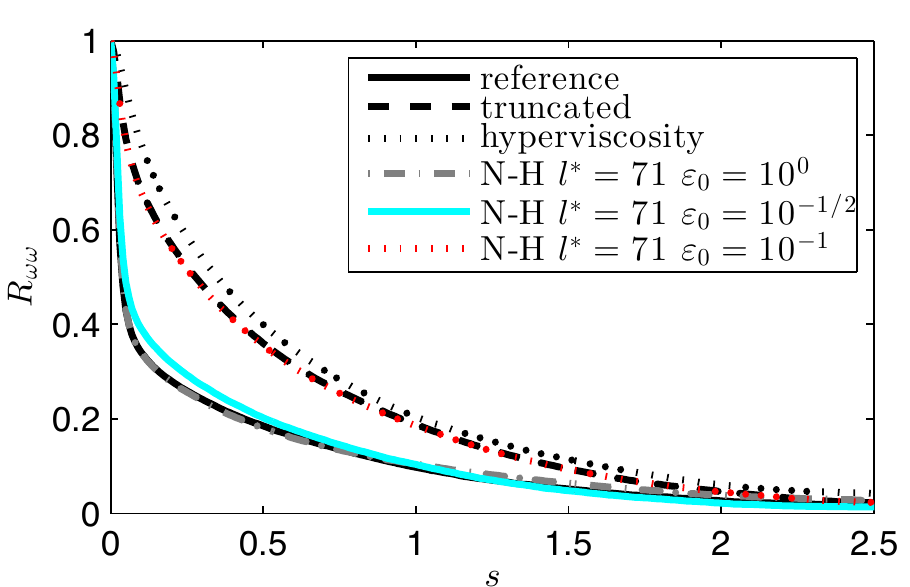}
    \includegraphics[width=.48\textwidth]{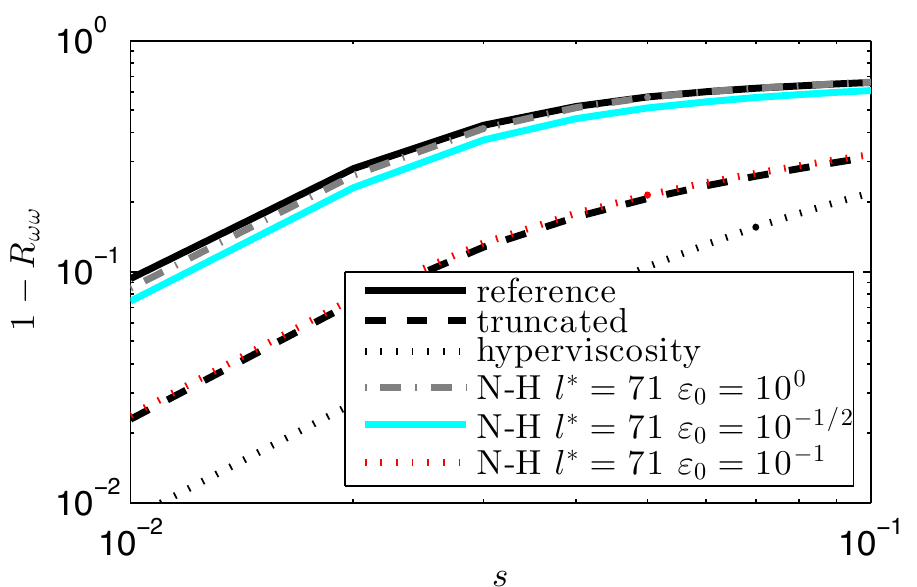}
  \end{center}%
  \caption{Autocorrelation functions $R_{\omega\omega}$ with small scale forcing. The Nos\'{e}-Hoover thermostat using various choices of the perturbation parameter $\varepsilon_0$ is compared to the reference solution (solid black), a truncated simulation (dashed black) and a hyperviscous model (dotted black).  The figure on the right shows $1-R_{\omega\omega}$ on a double logarithmic plot, to focus on short time scale behaviour.}
  \label{fig:acf_NH_mus_2s}
\end{figure}

\subsection{Ensemble dispersion}
Again we compare the evolution of the phase angle of the (0,1)-mode in a 50-member ensemble simulation for each of the three models. All simulations are performed with identical realizations of the Wiener process for the large scale forcing. Figure \ref{fig:ensemble_2s} compares the reference (center) to the hyperviscous (left) and Nos\'e-Hoover control (right) methods. The reference and Nos\'{e}-Hoover controlled ensembles show significant decorrelation at about time $t=5$, whereas for the model with hyperviscosity, the decorrelation is delayed until time $t=7$ or $t=8$. 
In the reference there is a notable split of the ensembles around $t=5$ into two main branches. This split can also be observed in the Nos\'e-Hoover approach, but not in the hyperviscosity model. At the final time $t=10$, both reduced models are slightly underdispersive.
\begin{figure}
  \begin{center}
    \includegraphics[width=\textwidth]{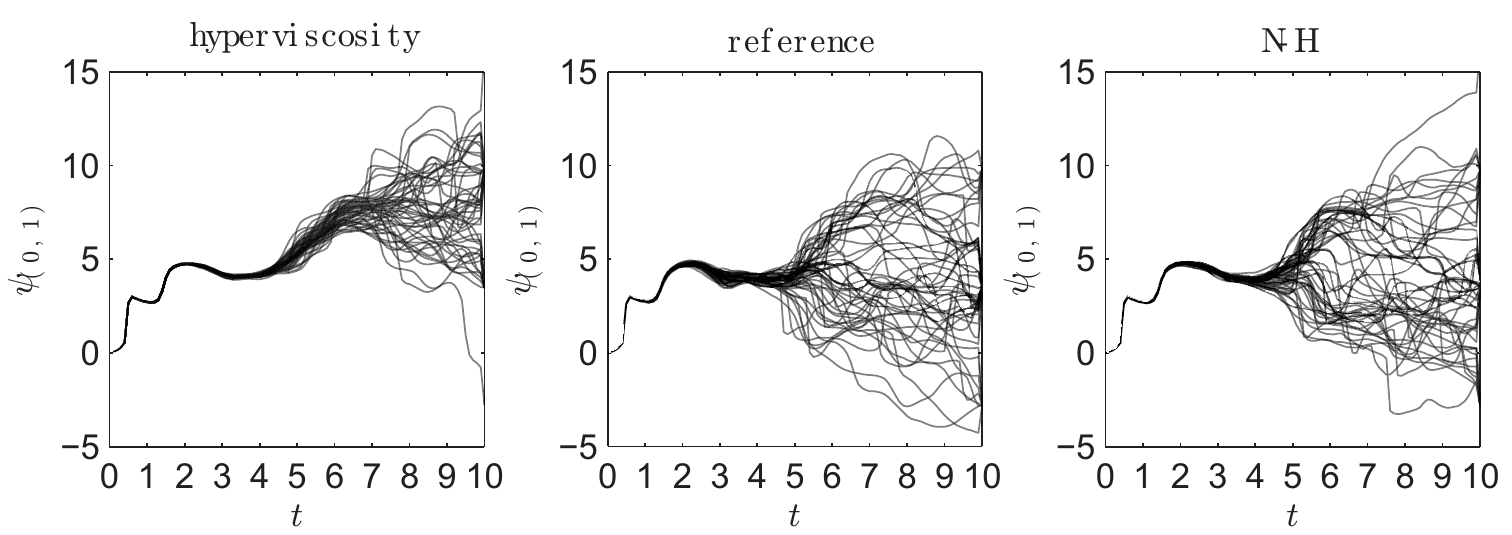}
  \end{center}%
  \caption{Ensemble dispersion in a simulation with small scale forcing. Plotted is the phase angle of the (0,1)-mode using (left) the model with artificial viscosity, (centre) the reference simulation, and (right) the Nos\'{e}-Hoover as indicated with $\dagger$ in table \ref{tab:pars}.}
  \label{fig:ensemble_2s}
\end{figure}

\section{Discussion}
\label{sec:discussion}
We have shown that the Nos\'{e}-Hoover method can be used to enforce a target background kinetic energy spectrum in 2D turbulence models with stochastic forcing, even when truncated well below the viscous scale. The parameterization comes at the mild cost of one additional dynamic variable for each energy shells controlled. 

In the experiments reported in the previous section, the target spectrum was inferred from a high resolution simulation, but it is important to emphasize that the target spectrum could also be taken from experiments or theory.  In particular, the method described here could be developed to enforce a $k^{-3}$ spectrum in low resolution models, if so desired.  The approach makes no explicit use of two-dimensional structure and hence is potentially extensible to 3D turbulence.  

\citet{FrSc:99} simulate two-dimensional turbulent dynamics using a wavelet basis. In this setting applying our control method might be even more effective, as the action can be restricted to the homogenous regions away from coherent structures.

\citet{SuGaCh:99} observe that the large scale dissipation influences the spectral slope in the forward enstrophy cascade. This means that it might be useful to apply a control on this end of the spectrum to perform simulations that display the hypothsized $k^{-3}$ spectrum in the high wavenumbers.\\

The authors express gratitude to Aleksandar Donev and Jacques Vanneste for very useful comments on earlier drafts of this paper.  The work of the second author was supported by grant  EP/G036136/1 from the Engineering and Physical Sciences Research Council (UK). The third author was supported by a grant financed by the Netherlands Organisation for Scientific Research (NWO), as part of research programme 613.001.009.

\appendix
\section{Forcing}
\label{sec:forcing}
With the energy given by $E(t) = -\frac{1}{2} \sum_{\bs{k}} \Delta^{-1}_{\bs{k}} \omega_{\bs{k}} \omega_{\bs{k}}^*$,  $\Delta^{-1}_{\bs{k}} = -|\bs{k}|^{-2}$, the expected power input due to forcing is equal to the expected change in energy (using It\^o's formula)
\begin{equation}
  \expbra{\dd E} 
= \expbra{\pd{E}{t} \dd t + \sum_{\bs{k}} \pd{E}{\omega_{\bs{k}}} \dd \omega_{\bs{k}} + \frac{1}{2} \dd \omega_{\bs{k}} \pds{E}{\omega_{\bs{k}}} \dd \omega_{\bs{k}}}.  \label{dE}
\end{equation}
For forcing with uniform magnitude across a band ${\bs{k}}\in K_f$ of wave numbers, 
we substitute $\dd \omega_{\bs{k}} = \hat f \dd B_{\bs{k}}$ into \eqref{dE} to find
\begin{align}
  \expbra{\dd E} 
&= -\hat{f}^2 \sum_{{\bs{k}} \in K_f} \Delta_{\bs{k}}^{-1} \dd t = P \dd t,
\end{align}
whence it follows that the magnitude of the forcing in the forced modes should be equal to $\hat{f} = \sqrt{\frac{P}{\sum_{{\bs{k}} \in K_f} |\bs{k}|^{-2}}}$. In the splitting method used, there are two forcing time steps for a time step of $\Delta t$. Each has the following implementation.
\begin{equation}
  \omega^+_{\bs{k}} = \omega_{\bs{k}} + \sqrt{\frac{P \Delta t/2}{\sum_{{\bs{k}} \in K_f} |\bs{k}|^{-2}}} (R_{\bs{k}} + \ui S_{\bs{k}}),
\end{equation}
where $R_{\bs{k}}, S_{\bs{k}}$ are unit normal pseudorandom numbers

\bibliographystyle{jfm}

\end{document}

%% file: macros.tex
\let\oldsqrt\sqrt
\def\sqrt{\mathpalette\DHLhksqrt}
\def\DHLhksqrt#1#2{%
\setbox0=\hbox{$#1\oldsqrt{#2\,}$}\dimen0=\ht0
\advance\dimen0-0.2\ht0
\setbox2=\hbox{\vrule height\ht0 depth -\dimen0}%
{\box0\lower0.4pt\box2}}
\renewcommand{\epsilon}{\varepsilon}

\makeatletter
\def\psls@dotteddashed{%
[ 0 \psk@dotsep CLW add \psk@dash ] 0 setdash  1 setlinecap  stroke}
\makeatother

\newcommand{\dd}{\,\mathrm{d}}

\newcommand{\bs}[1]{\boldsymbol{#1}}

\newcommand{\pd}[2]{\frac{\partial #1}{\partial #2}}

\newcommand{\pds}[2]{\frac{\partial^2 #1}{\partial #2^2}}

\newcommand{\ensbra}[1]{\left\langle #1 \right\rangle}
\newcommand{\expbra}[1]{\mathbb{E} \left[ #1 \right]}

\newcommand{\half}{\frac{1}{2}}


%% file: spectrum-2014-12-11_jfm.bbl
\begin{thebibliography}{52}
\expandafter\ifx\csname natexlab\endcsname\relax\def\natexlab#1{#1}\fi

\bibitem[Abramov \& Majda(2003)]{AbMa:03}
{\sc Abramov, R.~V. \& Majda, A.~J.} 2003 Statistically relevant conserved
  quantities for truncated quasigeostrophic flow. {\em Proceedings of the
  National Academy of Sciences\/} {\bf 100}~(7), 3841--3846.

\bibitem[Batchelor(1969)]{Batchelor:69}
{\sc Batchelor, G.~K.} 1969 Computation of the energy spectrum in homogeneous
  two-dimensional turbulence. {\em Physics of Fluids (1958-1988)\/} {\bf
  12}~(12), II--233.

\bibitem[Berner {\em et~al.\/}(2009)Berner, Shutts, Leutbecher \&
  Palmer]{Berner:09}
{\sc Berner, J., Shutts, G.~J., Leutbecher, M. \& Palmer, T.~N.} 2009 A
  spectral stochastic kinetic energy backscatter scheme and its impact on
  flow-dependent predictability in the ecmwf ensemble prediction system. {\em
  Journal of the Atmospheric Sciences\/} {\bf 66}~(3), 603--626.

\bibitem[Bofetta \& Musacchio(2010)]{Bofetta:10}
{\sc Bofetta, G. \& Musacchio, S.} 2010 Evidence for the double cascade
  scenario in two-dimensional turbulence. {\em Physical Review E\/} {\bf 82}.

\bibitem[Bouchet \& Venaille(2012)]{BoVe12}
{\sc Bouchet, F. \& Venaille, A.} 2012 Statistical mechanics of two-dimensional
  and geophysical flows. {\em Physics Reports\/} {\bf 515}~(5), 227--295.

\bibitem[Bracco {\em et~al.\/}(2000)Bracco, McWilliams, Murante, Provenzale \&
  Weiss]{BrMcMuPrWe:00}
{\sc Bracco, A., McWilliams, J., Murante, G., Provenzale, A. \& Weiss, J.} 2000
  Revisiting freely decaying two-dimensional turbulence at millennial
  resolution. {\em Physics of Fluids (1994-present)\/} {\bf 12}~(11),
  2931--2941.

\bibitem[Canuto {\em et~al.\/}(2006)Canuto, Hussaini, Quarteroni \&
  Zang]{Canuto:06}
{\sc Canuto, C., Hussaini, M.~Y., Quarteroni, A. \& Zang, T.~A.} 2006 {\em
  Spectral Methods: Fundamentals in Single Domains\/}. Springer Verlag.

\bibitem[Danilov \& Gurarie(2000)]{DaGu:00}
{\sc Danilov, S.~D. \& Gurarie, D.} 2000 Quasi-two-dimensional turbulence. {\em
  Physics-Uspekhi\/} {\bf 43}~(9), 863.

\bibitem[Delong {\em et~al.\/}(2013)Delong, Griffith, Vanden-Eijnden \&
  Donev]{DeGrVaDo13}
{\sc Delong, S., Griffith, B.~E., Vanden-Eijnden, E. \& Donev, A.} 2013
  Temporal integrators for fluctuating hydrodynamics. {\em Physical Review E\/}
  {\bf 87}~(3), 033302.

\bibitem[Domaradzki \& Saiki(1997)]{DoSa:97}
{\sc Domaradzki, J.~A. \& Saiki, E.~M.} 1997 Backscatter models for large-eddy
  simulations. {\em Theoretical and computational fluid dynamics\/} {\bf
  9}~(2), 75--83.

\bibitem[Donev {\em et~al.\/}(2010)Donev, Vanden-Eijnden, Garcia \&
  Bell]{DoVaGaBe10}
{\sc Donev, A., Vanden-Eijnden, E., Garcia, A. \& Bell, J.} 2010 On the
  accuracy of finite-volume schemes for fluctuating hydrodynamics. {\em
  Communications in Applied Mathematics and Computational Science\/} {\bf
  5}~(2), 149--197.

\bibitem[Dubinkina \& Frank(2007)]{DuFr:07}
{\sc Dubinkina, S. \& Frank, J.} 2007 Statistical mechanics of {A}rakawa's
  discretizations. {\em Journal of Computational Physics\/} {\bf 227}~(2),
  1286--1305.

\bibitem[Dubinkina \& Frank(2010)]{DuFr:10}
{\sc Dubinkina, S. \& Frank, J.} 2010 Statistical relevance of vorticity
  conservation in the hamiltonian particle-mesh method. {\em Journal of
  Computational Physics\/} {\bf 229}~(7), 2634--2648.

\bibitem[E {\em et~al.\/}(2001)E, Mattingly \& Sinai]{EmaSi01}
{\sc E, W., Mattingly, J. \& Sinai, Y.} 2001 Gibbsian dynamics and ergodicity
  for the stochastically forced navier-stokes equation. {\em Communications in
  Mathematical Physics\/} {\bf 224}~(1), 83--106.

\bibitem[E \& Mattingly(2001)]{EMa:01}
{\sc E, W. \& Mattingly, J.~C.} 2001 Ergodicity for the navier-stokes equation
  with degenerate random forcing: Finite-dimensional approximation. {\em
  Communications on Pure and Applied Mathematics\/} {\bf 54}~(11), 1386--1402.

\bibitem[Evensen(2009)]{Evensen:09}
{\sc Evensen, G.} 2009 {\em Data Assimilation: the Ensemble Kalman Filter\/},
  2nd edn. Springer.

\bibitem[Farazmand {\em et~al.\/}(2011)Farazmand, Kevlahan \&
  Protas]{FaKePr:11}
{\sc Farazmand, M., Kevlahan, N.-R. \& Protas, B.} 2011 Controlling the dual
  cascade of two-dimensional turbulence. {\em Journal of Fluid Mechanics\/}
  {\bf 668}, 202--222.

\bibitem[Farge {\em et~al.\/}(1996)Farge, Kevlahan, Perrier \&
  Goirand]{FaKePeGo:96}
{\sc Farge, M., Kevlahan, N., Perrier, V. \& Goirand, E.} 1996 Wavelets and
  turbulence. {\em Proceedings of the IEEE\/} {\bf 84}~(4), 639--669.

\bibitem[Fatkullin \& Vanden-Eijnden(2004)]{FaVa04}
{\sc Fatkullin, I. \& Vanden-Eijnden, E.} 2004 A computational strategy for
  multiscale systems with applications to lorenz 96 model. {\em Journal of
  Computational Physics\/} {\bf 200}~(2), 605--638.

\bibitem[Frank \& Gottwald(2011)]{FrGo11}
{\sc Frank, J. \& Gottwald, G.} 2011 The {L}angevin limit of the
  {N}os\'{e}-{H}oover-{L}angevin thermostat. {\em Journal of Statistical
  Physics\/} {\bf 143}~(4), 715--724.

\bibitem[Frisch(1995)]{Frisch:95}
{\sc Frisch, U.} 1995 {\em Turbulence: the legacy of {A.N.} {Kolmogorov}\/}.
  Cambridge University Press.

\bibitem[Fr{\"o}hlich \& Schneider(1999)]{FrSc:99}
{\sc Fr{\"o}hlich, J. \& Schneider, K.} 1999 Computation of decaying turbulence
  in an adaptive wavelet basis. {\em Physica D: Nonlinear Phenomena\/} {\bf
  134}~(3), 337--361.

\bibitem[Gotoh(1998)]{Gotoh:98}
{\sc Gotoh, T.} 1998 Energy spectrum in the inertial and dissipation ranges of
  two-dimensional steady turbulence. {\em Physical Review E\/} {\bf 57}~(3),
  2984.

\bibitem[Grooms \& Majda(2013)]{GrMa2013}
{\sc Grooms, I. \& Majda, A.~J.} 2013 Efficient stochastic
  superparameterization for geophysical turbulence. {\em Proceedings of the
  National Academy of Sciences\/} {\bf 110}~(12), 4464--4469.

\bibitem[Hairer {\em et~al.\/}(2006)Hairer, Lubich \& Wanner]{HaLuWa:06}
{\sc Hairer, E., Lubich, C. \& Wanner, G.} 2006 {\em Geometric numerical
  integration: structure-preserving algorithms for ordinary differential
  equations\/}, , vol.~31. Springer.

\bibitem[Hasselmann(1976)]{Hasselman76}
{\sc Hasselmann, K.} 1976 Stochastic climate models part i. theory. {\em
  Tellus\/} {\bf 28}~(6), 473--485.

\bibitem[Hoover(1985)]{Hoover85}
{\sc Hoover, W.~G.} 1985 Canonical dynamics: equilibrium phase-space
  distributions. {\em Physical Review A\/} {\bf 31}~(3), 1695.

\bibitem[Jones \& Leimkuhler(2011)]{JoLe11}
{\sc Jones, A. \& Leimkuhler, B.} 2011 Adaptive stochastic methods for sampling
  driven molecular systems. {\em The Journal of Chemical Physics\/} {\bf
  135}~(8), --.

\bibitem[Kent {\em et~al.\/}(2012)Kent, Thuburn \& Wood]{KeThWo:12}
{\sc Kent, J., Thuburn, J. \& Wood, N.} 2012 Assessing implicit large eddy
  simulation for two-dimensional flow. {\em Quarterly Journal of the Royal
  Meteorological Society\/} {\bf 138}~(663), 365--376.

\bibitem[Kraichnan(1967)]{Kraichnan:67}
{\sc Kraichnan, R.~H.} 1967 Inertial ranges in two-dimensional turbulence. {\em
  Physics of Fluids\/} {\bf 10}, 1417--1423.

\bibitem[Kraichnan(1971)]{Kraichnan:71}
{\sc Kraichnan, R.~H.} 1971 Inertial-range transfer in two-and
  three-dimensional turbulence. {\em Journal of Fluid Mechanics\/} {\bf
  47}~(03), 525--535.

\bibitem[Leimkuhler {\em et~al.\/}(2009)Leimkuhler, Noorizadeh \&
  Theil]{Leimkuhler:09}
{\sc Leimkuhler, B., Noorizadeh, E. \& Theil, F.} 2009 A gentle thermostat for
  molecular dynamics. {\em J. Stat. Phys.\/} {\bf 135}~(2), 261--277.

\bibitem[Leimkuhler \& Reich(2004)]{Leimkuhler:04}
{\sc Leimkuhler, B. \& Reich, S.} 2004 {\em Simulating Hamiltonian Dynamics\/},
  1st edn. {\em Camb. Monogr. on Appl. and Comput. Math.\/} . Camb., U.K.:
  Camb. Univ. Press.

\bibitem[Leith(1968)]{Leith:68}
{\sc Leith, C.~E.} 1968 Diffusion approximation for two-dimensional turbulence.
  {\em Physics of Fluids (1958-1988)\/} {\bf 11}~(3), 671--672.

\bibitem[Lindborg(1999)]{Lindborg:99}
{\sc Lindborg, E.} 1999 Can the atmospheric kinetic energy spectrum be
  explained by two-dimensional turbulence? {\em Journal of Fluid Mechanics\/}
  {\bf 388}, 259--288.

\bibitem[Majda {\em et~al.\/}(2001)Majda, Timofeyev \&
  Vanden~Eijnden]{MaTiVa01}
{\sc Majda, A.~J., Timofeyev, I. \& Vanden~Eijnden, E.} 2001 A mathematical
  framework for stochastic climate models. {\em Communications on Pure and
  Applied Mathematics\/} {\bf 54}~(8), 891--974.

\bibitem[Miller(1990)]{Miller90}
{\sc Miller, J.} 1990 Statistical mechanics of euler equations in two
  dimensions. {\em Physical Review Letters\/} {\bf 65}~(17), 2137.

\bibitem[Mori(1965)]{Mori65}
{\sc Mori, H.} 1965 Transport, collective motion, and brownian motion. {\em
  Progress of Theoretical Physics\/} {\bf 33}~(3), 423--455.

\bibitem[Nastrom \& Gage(1985)]{NaGa:85}
{\sc Nastrom, G. \& Gage, K.~S.} 1985 A climatology of atmospheric wavenumber
  spectra of wind and temperature observed by commercial aircraft. {\em Journal
  of the Atmospheric Sciences\/} {\bf 42}~(9), 950--960.

\bibitem[Nastrom {\em et~al.\/}(1984)Nastrom, Gage \& Jasperson]{Nastrom:84}
{\sc Nastrom, G.~D., Gage, K.~S. \& Jasperson, W.~H.} 1984 Kinetic energy
  spectrum of large-and mesoscale atmospheric processes. {\em Nature\/} {\bf
  310}, 36--38.

\bibitem[Nos\'{e}(1984{\natexlab{{\em a\/}}})]{Nose84a}
{\sc Nos\'{e}, S.} 1984{\natexlab{{\em a\/}}} A molecular dynamics method for
  simulations in the canonical ensemble. {\em Molecular physics\/} {\bf
  52}~(2), 255--268.

\bibitem[Nos\'{e}(1984{\natexlab{{\em b\/}}})]{Nose84b}
{\sc Nos\'{e}, S.} 1984{\natexlab{{\em b\/}}} A unified formulation of the
  constant temperature molecular dynamics methods. {\em The Journal of Chemical
  Physics\/} {\bf 81}~(1), 511--519.

\bibitem[Robert(1991)]{Robert91}
{\sc Robert, R.} 1991 A maximum-entropy principle for two-dimensional perfect
  fluid dynamics. {\em Journal of Statistical Physics\/} {\bf 65}~(3-4),
  531--553.

\bibitem[Robert \& Sommeria(1991)]{RoSo91}
{\sc Robert, R. \& Sommeria, J.} 1991 Statistical equilibrium states for
  two-dimensional flows. {\em Journal of Fluid Mechanics\/} {\bf 229},
  291--310.

\bibitem[Saffman(1971)]{Saffman:71}
{\sc Saffman, P.} 1971 On the spectrum and decay of random two-dimensional
  vorticity distributions at large {R}eynolds number. {\em Studies in Applied
  Mathematics\/} {\bf 50}, 377--383.

\bibitem[Samoletov {\em et~al.\/}(2007)Samoletov, Dettmann \&
  Chaplain]{SaDeCh07}
{\sc Samoletov, A.~A., Dettmann, C.~P. \& Chaplain, M.~A.} 2007 Thermostats for
  ``slow'' configurational modes. {\em Journal of Statistical Physics\/} {\bf
  128}~(6), 1321--1336.

\bibitem[Shutts(2005)]{Shutts:05}
{\sc Shutts, G.} 2005 A kinetic energy backscatter algorithm for use in
  ensemble prediction systems. {\em Quarterly Journal of the Royal
  Meteorological Society\/} {\bf 131}~(612), 3079--3102.

\bibitem[Sukoriansky {\em et~al.\/}(1999)Sukoriansky, Galperin \&
  Chekhlov]{SuGaCh:99}
{\sc Sukoriansky, S., Galperin, B. \& Chekhlov, A.} 1999 Large scale drag
  representation in simulations of two-dimensional turbulence. {\em Physics of
  Fluids (1994-present)\/} {\bf 11}~(10), 3043--3053.

\bibitem[Thuburn {\em et~al.\/}(2014)Thuburn, Kent \& Wood]{ThKeWo:14}
{\sc Thuburn, J., Kent, J. \& Wood, N.} 2014 Cascades, backscatter and
  conservation in numerical models of two-dimensional turbulence. {\em
  Quarterly Journal of the Royal Meteorological Society\/} {\bf 140}~(679),
  626--638.

\bibitem[Tulloch \& Smith(2006)]{TuSm06}
{\sc Tulloch, R. \& Smith, K.} 2006 A theory for the atmospheric energy
  spectrum: Depth-limited temperature anomalies at the tropopause. {\em
  Proceedings of the National Academy of Sciences\/} {\bf 103}~(40),
  14690--14694.

\bibitem[Xing {\em et~al.\/}(2009)Xing, Majda \& Grabowski]{XiMaGr09}
{\sc Xing, Y., Majda, A.~J. \& Grabowski, W.~W.} 2009 New efficient sparse
  space-time algorithms for superparameterization on mesoscales. {\em Monthly
  Weather Review\/} {\bf 137}~(12), 4307--4324.

\bibitem[Zwanzig(1961)]{Zwanzig61}
{\sc Zwanzig, R.} 1961 Memory effects in irreversible thermodynamics. {\em
  Physical Review\/} {\bf 124}~(4), 983.

\end{thebibliography}
